\documentclass[12pt]{article}
\usepackage{amssymb,amsmath}
\usepackage{epsfig}
\newcommand{\sect}[1]{\setcounter{equation}{0}\section{#1}}

\def\L{{\mathcal L}}

\def\J{{\mathcal J}}

\def\a{\alpha}

\def\g{\gamma}

\def\s{\sigma}
\def\t{\tau}

\def\k{\kappa}

\def\e{\eta}

\def\rb{\right}
\def\lb{\left}

\newcommand{\eq}[1]{\begin{equation} #1 \end{equation}}
\newcommand{\al}[1]{\begin{align} #1 \end{align}}
\newcommand{\ml}[1]{\begin{multline} #1 \end{multline}}

\begin{document}
\begin{center}
{\bf{\Large Generalized pulsating strings} \\
\vspace*{.35cm}
}

\vspace*{1cm}
H. Dimov${}^{\ddag}$, 
R.C. Rashkov${}^{\dagger}$\footnote{e-mail: rash@phys.uni-sofia.bg}

\ \\
${}^{\dagger}$Department of Physics, Sofia University, 1164 Sofia,
Bulgaria 

\ \\

${}^{\ddag}$ Department of Mathematics, University of 
Chemical Technology and Metallurgy, 1756 Sofia, Bulgaria

\end{center}

\vspace*{.8cm}

\begin{abstract}
In this paper we consider new solutions for pulsating strings. For
this purpose we use the idea of the generalized ansatz for folded 
and circular strings in hep-th/0311004. We find the
solutions to the resulting Neumann-Rosochatius integralbe system and
the corrections to the energy. To do that we use the approach
developed by Minahan in hep-th/0209047 and find that the corrections 
are quite different from
those obtained in that paper and in hep-th/0310188. We conclude with
comments on our solutions and the obtained corrections to the energy,
expanded to the leading order in lambda.
\end{abstract}

\vspace*{.8cm}

\sect{Introduction}

The recent developments in string theory is basically related to
searching more and more examples supporting and verifying AdS/CFT
correspondence. After a huge amount of papers checking the conjecture
at supergravity level, a new approaches treating the problem beyond
the SUGRA limit appeared. First of all, the significan development in
our understanding of string theory in $AdS_5\times S^5$ background
was inspired by the idea of Berenstein, Maldacena and Nastase (BMN) to
analyse  the string theory in the visinity of null geodesics of
$AdS_5\times S^5$ space \cite{bmn}. Expanding string sigma model around the
geodesics one ends up with so called pp-wave geometry and procedure
tunrs out to be so called Penrose limit. The remarkable properties of
this background are that it has a maximal amount of supersymmetry and
the string sigma model in this background turns out to be exactly
solvable. Putting all these together, one can conclude that the
resulting theory is semiclassical string theory of point-like string
moving along the geodesics and having large angular momentum $J$.

The picture emerging from BMN limit was understood by using another
idea - semiclassical analysis of string theory in $AdS_5\times S^5$
background \cite{gkp}. The key point in this approach is that large classical
charges means small $\sigma$-model corrections so that the
semiclassical results are reliable enough to compute the dimensions of
the operators from SYM side. This approach enables us to consider more
generic operators than just BPS ones. In a considerable amount of
papers
\cite{ts,ts1,ts2,ts3,rash,rv,mz,zar,tsstau,krucz,tsneum,tsnr,frolts,rash1,
lar,herlop,ryang,mosaffa1,mosaffa2,mosaffa3,krist}\footnote{For more complete list see the
  references in these papers.} various string 
configurations were analysed and the
anomalous dimensions of the corresponding SYM operators were
obtained. One should note that most of the examples of string
configurations corresponding to generically non-BPS operators share
one property - in the large energy limit all they become
supersymmetric \cite{towns1, towns2}. One of the exeptions from the so
called ´´nearly''BPS strings are configurations named pulsating strings
\cite{min,minzar}. The corresponding operators also have large
charges, but are not necessarily close to any BPS one.

In this paper we are looking for more general solutiuons belonging to
the class of pulsating strings. Before we go into details, let us
remind some of the general properties of the pulsating strings. Let us
consider pulsating string in $S^5$ part of $AdS_5\times S^5$
geometry. The string states then should be characterized by the
charges associated to the isometry group of $S^5$, $SO(6)$. The
highest weight representation $(J_1,J_2,J_3)$ is actually defined by
these angular momenta. On gauge theory side we know that SYM theory
can be characterized by three $R$-charges $(J_1,J_2,J_3)$ which define
the highes weight representation of $SO(6)$ gauge group with Dynkin indices
$[J_2+J_3,J_1-J_2,J_2-J_3]$ \cite{ts2}. The operators in SYM
corresponding to the above $R$-charges are of the form
\eq{
Tr\lb(X^{J_1}Y^{J_2}Z^{J_3}\rb)
\label{i1}
}
where $X,Y$ and $Z$ are three complex scalars in the adjoint of
$SO(6)$ forming $N=4$ supermultiplet. Due to the mixing between the
operators, the computation of the anomalous dimensions becomes a
non-trivial problem even at one loop. Going back, one should note that
for different string configurations known in the literature there are
more ot less complicated relations between the anomalous dimensions
and the $R$-charges. As we already mentioned above, the pulsating
strings are one of the examples which has more special status. The
goal of this paper is to generalize the known solutions for pulsating
strings \cite{min,minzar} and to find the corrections to the classical
energy, which means to obtain the anomalous dimensions. The paper is
organized as follows. In the next section we briefly review the known
in the literature solutions. In Section 3, based on the generalized
string ansatz for folded and circular string of \cite{tsnr}\footnote{A
  similar ansatz was suggested also in \cite{plamenb} and \cite{plamenb1}.} we
find new solutions for pulsating strings. In the Section 4 we find
the corrections to the classical energy for our solutions and find the
anomalous dimension of the corresponding SYM operators.

\sect{Review of pulsating strings}

In this section we give a brief review of the pulsating string
solutions obtained first by Minahan \cite{min} and generalized later
in \cite{minzar}. We will concentrate only on the case of pulsating
string on $S^5$ part of $AdS_5\times S^5$, i.e. we consider a circular
string which pulsates expanding and contracting on $S^5$. To fix the
notations, we start with the metric of $S^5$ and relevant part of
$AdS_5$
\eq{
ds^2=R^2\lb(\cos^2\theta d\Omega_3^2+d\theta^2+\sin^2\theta d\psi^2+
d\rho^2-cosh^2 dt^2\rb),
\label{1.1}
}
where $R^2=2\pi{\a}^\prime\sqrt{\lambda}$. To obtain the simplest
solution, we identify the target space time with the worldsheet
one, $t=\tau$, and use the ansatz $\psi=m\sigma$, i.e. the string is
stretched along $\psi$ direction, $\theta=\theta(\tau)$ and keep the
dependence on $\rho=\rho(\tau)$ for a while. The reduced Nambu-Goto
action in this case is
\eq{
S=m\sqrt{\lambda}\int dt \sin\theta \sqrt{cosh^2\rho-\dot\theta^2}.
\label{1.1a}
}
For our considerations it is useful to pass to Hamiltonian
formulation. For this purpose we find the canonical momenta and find
the Hamiltonian in the form\footnote{For more details see \cite{min}.}
\eq{
H=cosh\rho\sqrt{\Pi_\rho^2+\Pi_\theta^2+m^2\lambda\sin^2\theta}.
\label{1.2}
}
Fixing the string to be at the origin of $AdS_5$ space ($\rho=0$), we
see that the squared Hamiltonian have a form very similar to a point
particle. The last term in (\ref{1.2}) can be considered as a
perturbation, so first we find the wave function for a free particle
in the above geometry
\ml{
\frac{cosh\rho}{sinh^3\rho}\dfrac{d}{d\rho}cosh\rho sinh^3\rho
\dfrac{d}{d\rho}\Psi(\rho,\theta)-\frac{cosh^2\rho}{\sin\theta
  \cos^3\theta} \dfrac{d}{d\theta}\sin\theta\,\cos^3\theta
\dfrac{d}{d\theta} \Psi(\rho,\theta)\\
=E^2\Psi(\rho,\theta).
\label{1.3}
}
The solution to the above equation is
\eq{
\Psi_{2n}(\rho,\theta)=(cosh\rho)^{-2n-4}\, P_{2n}(\cos\theta)
\label{1.4}
}
where $P_{2n}(\cos\theta)$ are spherical harmonics on $S^5$ and the
energy is given by
\eq{
E_{2n}=\Delta=2n+4.
\label{1.5}
}
Since we consider highly excited states (large energies), one should
take large $n$, so we can approximate the spherical harmonics as
\eq{
P_{2n}(\cos\theta)\approx \sqrt{\frac{4}{\pi}}\cos(2n\theta).
\label{1.6}
}
The correction to the energy can be obtained by using perturbation
theory, which to first order is
\eq{
\delta E^2=\int\limits_0^{\pi/2}d\theta\,
\Psi_{2n}^\star(0,\theta)\,m^2\lambda \sin^2\theta\,
\Psi_{2n}(\theta)
=\frac{m^2\lambda}{2}.
\label{1.7}
}
Up to first order in $\lambda$ we find for the anomalous dimension of
the corresponding YM operators\footnote{See \cite{min} for more
  details.}
\eq{
\Delta-4=2n[1+\frac 12\,\frac{m^2\lambda}{(2n)^2}].
\label{1.8}
}
On should note that in this case the $R$-charge is zero. In order to
include it, we consider pulsating string on $S^5$ which has a center
of mass that is moving on the $S^3$ subspace of $S^5$ \cite{minzar}. While in the
previous example $S^3$ part of the metric was assumed trivial, now we
consider all the $S^3$ angles to depend on $\tau$ (only). The
corresponding Nambu-Goto action now is
\eq{
S=-m\sqrt{\lambda}\int\,
dt\sin\theta\,\sqrt{1-\dot\theta^2-\cos^2\theta
  g_{ij}\dot\phi^i\dot\phi^j },
\label{1.9}
}
where $\phi_i$ are $S^3$ angles and $g_{ij}$ is the corresponding
$S^3$ metric. The Hamiltonian in this case is \cite{minzar}
\eq{
H=\sqrt{\Pi_\theta^2+\frac{g^{ij}\Pi_i\Pi_j}{\cos^2\theta}
  +m^2\lambda\sin^2\theta}.
\label{1.10}
}
Again, we see that the squared Hamiltonian looks like the point
particle one, however, now the potential has angular
dependence. Denoting the quantum number of $S^3$ and $S^5$ by $J$ and
$L$ correspondingly, one can write the Schrodinger equation 
\eq{
-\frac{4}{\omega}\dfrac{d}{d\omega}\Psi(\omega)
+\frac{J(J+1)}{\omega}\Psi(\omega) = L(L+4) \Psi(\omega),
\label{1.11}
}
where we set $\omega=\cos^2\theta$. The solution to the Schrodinger
equation is
\eq{
\Psi(\omega)=\frac{\sqrt{2(l+1)}}{(l-j)!}\,\frac{1}{\omega}\left(
  \frac{d}{d\omega}\right)^{l-j}
\omega^{l+j}(1-\omega)^{l-j}
, \quad j=\frac{J}{2}; l=\frac{L}{2}.
\label{1.12}
}
The first order correction to the energy $\delta E$ is
\eq{
\delta E^2=m^2\lambda\,\frac{2(l+1)^2-(j+1)^2-j^2}{(2l+1)(2l+3)},
\label{1.13}
}
or, up to first order in $\lambda$
\eq{
E^2=L(L+4)+m^2\lambda\frac{L^2-J^2}{2L^2}
\label{1.14}
}
The anomalous dimension then is given by
\eq{
\gamma=\frac{m^2\lambda}{4L}\a(2-\a),
\label{1.15}
}
where $\a=1-J/L$.

We conclude this section reffering for more details to \cite{min} and
\cite{minzar}.

\sect{New Pulsating string solutions}

In this section we consider in details generalized solutions
generated by the more general ansatz suggested in \cite{tsnr}. After a
short review of the generalized ansatz we will analyse in some details
the solution of the resulting dynamical system. 

First of all, we will consider the puslating string solutions in $S^5$
part of $AdS_5\times S^5$ geometry of the target spacetime. To start
with, let us mention the generalized ansatz for the string embedding
coordinates:
\al{
& t=\k\t \notag \\
& X_i=r_i(\t)e^{i\a(\t)+im_i\s}=z_i(\t)e^{m_i\s} \label{2.1} \\
& \sum\limits_{i=1}^3r_i^2(\t)=1 \notag
}
and other coordinates in $AdS_5$ part set to appropriate constant
values. In order to ensure singlevelued parametrization of the string
worldsheet the constants $m_i$ should be integers. Next, to ensure
vanishing of the non-Cartan components of $O(6)$ angular momentum, one
should take $m_i\neq m_j$. No we are ready, using the ansatz
(\ref{2.1}), to write down the lagrangian of the constrained sigma
model describing our string dynamics:
\eq{
\L=\frac 12\sum\limits_{i=1}^3\lb(\dot z_i{\dot{\bar z_i}}-m_i^2z_i\bar
z_i\rb) \frac 12\Lambda\lb(\sum\limits_{i=1}^3z_i\bar z_i -1\rb)
\label{2.2}
}
where $z_i(\t)$ are defined in (\ref{2.1}), $\Lambda$ is a lagrange
multiplier and the coordinates satisfy the constraints
\eq{
z_i\bar z_i=r_i^2
\label{2.3}
}
(i.e., we are dealing with constrained $SO(6)$ sigma model). In terms
of the real variables ($r_i,\a_i$) the lagrangian can be written in
the form
\eq{
\L=\frac 12\sum\limits_{i=1}^3\lb[\dot
  r_i^2+r_i^2\dot\a^2-m_i^2r_i^2\rb]+\frac
12\Lambda\lb(\sum\limits_{i=1}^3 r_i^2-1\rb)
\label{2.4}
}
The equations of motion for $\a_i$ can be easily solved in terms of
$r_i$:
\eq{
\frac{d}{d\t}\lb(r_i^2\dot\a_i\rb)=0 \quad \Rightarrow \quad \dot\a=
\frac{\J_i}{r_i^2}
\label{2.5}
}
so that the lagrangian get easily reduced to the form
\eq{
\L=\frac 12\sum\limits_{i=1}^3\lb[\dot
  r_i^2-m_i^2r_i^2+\frac{\J_i}{r_i^2}\rb] +\frac 12\Lambda
\lb(\sum\limits_{i=1}^3r_i^2-1\rb)
\label{2.6}}
As in the case of folded and circular strings, one can consider two
cases - of constant lagrange multiplier $\Lambda$ and non-constant
one.

Let us focus first on the case of constant $\Lambda$. This case is the
simplest example of pulsating string solutions. If we use complex
notations, the lagrangian (\ref{2.2}) leads to a simple equation of
motion
\eq{
\ddot z_i+(m_i^2-\Lambda)z_i=0
\label{2.7}
}
If we denote $l_i^2=m_i-\Lambda$ and use the fact that $\Lambda$ is a
constant, the equation (\ref{2.7}) has the following general solution
\eq{
z_i=a_ie^{il_i\t}+b_ie^{-il_i\t}.
\label{2.8}
}
The assumption that $\Lambda$ is a constant implies that one can set
$b_i=0$ and the solution then becomes
\eq{
z_i=a_ie^{il_i\t}
\label{2.9}
}
supplied with the constraint
\eq{
\sum\limits_{i=1}^3a_i^2=1.
\label{2.10}
}

Now we turn to the case of non-constant lagrange multiplier
$\Lambda$. We will look for a solution of the constrained sigma model
choosing 
\al{
& v_i=0 \notag \\
& \a_i=0 \label{2.11} \\
& r_3=0.
}
i.e.
\eq{
x_i=r_i(\t)e^{im_i\s}, \quad \sum\limits_{i=1}^2r_i^2(\t)=1.
\label{2.12}
}
In this case the lagrangian and the lagrange multiplier become
\al{
& \L=\frac 12\sum\limits_{i=1}^2\lb(\dot r_i^2-m_i^2r_i^2\rb)+
\frac 12\Lambda\lb(\sum\limits_{i=1}^2r_i^2-1\rb) \label{2.13} \\
& \Lambda=-\sum\limits_{i=1}^2\lb(\dot r_i^2-m_i^2r_i^2\rb).
\label{2.14}
}
The equations of motion following from the above lagrangian are
\eq{
\ddot r_i=-m_i^2+\Lambda r_i.
\label{2.15}
}
We solve the equations using the natural parametrization of $S^3$ in
terms of the angles $\g$ and $\psi$
\al{
& r_1=\sin\g\cos\psi \notag \\
& r_2=\sin\g\sin\psi \label{2.16} \\
& r_3=\cos\g. \notag
}
The choice (\ref{2.11}) means that we set $\g=\pi/2$. The resulting
equation for the only dinamical variable $\psi$ becomes
\eq{
\ddot\psi+m_{21}^2\sin\psi\cos\psi=0
\label{2.17}
}
where $m_{21}^2=m_2^2-m_1^2$. Assuming that $\psi(0)=0$, we find
\eq{
A\t=\int\limits_0\psi\frac{d\psi}{\sqrt{1-\frac{m_{21}^2}{A}\sin^2\psi}},
\label{2.18}
}
where $A$ is an integration constant. In what follows we will
consider the case of
\eq{
\lb(\frac{m_{21}^2}{A}\rb)^2\leq 1
\label{2.19}
}
and the opposite case can be considered in the same way (we note that
the condition (\ref{2.19}) ensures the existing of ¨turning¨ point,
i.e. the string is pulsating). The solution for $\psi$ then is
\eq{
\sin\psi=sn\lb(A\t,t^2\rb)
\label{2.20}
}
where the modulus $t^2$ is defined as $t^2=m_{21}^2/A^2$. If we choose
the period of pulsations to be $2\pi$, the integration constant is
determined by this condition to
\eq{
A=\frac{2K(t)}{\pi},
\label{2.21}
}
where $K(t)$ is the complete elliptic integral of first kind.

Now we are to give essential generalization of the pulsating string
solution presented above. The case we will consider is of non-constant 
$\Lambda$ but with
\eq{
v_3=0, \quad \a_3=0, \quad r_3=0.
\label{2.22}
}
Using the parametrization (\ref{2.1}) and the constraints (\ref{2.22})
one can find for our pulsating string the following lagrangian
\eq{
\L=\frac 12\sum\limits_{i=1}^2\lb(\dot
r_i^2-m_i^2r_i^2+\frac{\J_i^2}{r_i^2}\rb)
+ \frac 12\Lambda\lb(\sum\limits_{i=1}^2 r_i^2-1\rb),
\label{2.23}
}
where we substituted for $\dot\a=\J_i^2/r_i^2$ (which follows from the
equation for $\a$). Solving for the lagrange multiplier $\Lambda$ we
find
\eq{
\Lambda=-\sum\limits_{i=1}^2\lb(\dot
r_i^2-m_i^2r_i^2+\frac{\J_i^2}{r_i^2} \rb).
\label{2.24}
}
The equations of motion are
\eq{
\ddot r_i=-m_i^2r_i-\frac{\J_i^2}{r_i^2}-r_i\lb(\sum\limits_{j=1}^2
\dot r_j^2-m_j^2r_j^2+\frac{\J_i^2}{r_j^2}\rb).
\label{2.25}
}
Using the explicit parametrization (\ref{2.1}) one can find the
following equation for the angle $\psi$
\eq{
\ddot\psi+m_{21}^2\sin\psi\cos\psi+\frac {\J_2^2\cos\psi}{\sin^3\psi} 
-\frac{\J_1^2\sin\psi}{\cos^3\psi}=0.
\label{2.26}
}
Integrating once the above equation one finds
\eq{
\dot\psi=\frac{\sqrt{\J_2^2+(A+\J_1^2+\J_2^2)\sin^2\psi-(A+m_{21}^2)
\sin^4\psi +m_{21}^2\sin^6\psi}}{\sin\psi\cos\psi}.
\label{2.27}
}
Defining $X=\sin^2\psi/2$ we get
\eq{
\dfrac{dX}{d\t}=\sqrt{\J_2^2+aX-bX^2 +8m_{21}^2X^3},
\label{2.28}
}
where
\al{
& a=2(A+\J_1^2+\J_2^2) \notag \\
& b=4(A+m_{21}^2).
\label{2.29}
}
Obviously this equation has a solution in terms of Jacobi elliptic functions.
To bring the equation (\ref{2.28}) into Jacobi canonical form we use
the change of variables
\eq{
\a+\xi.
\label{2.30}
}
If we denote by $\a$ the root of the equation
\eq{
8m_{21}^2\a^3-b\a^2+a\a+\J_2^2=0,
\label{2.31}
}
one can write the equation (\ref{2.28}) in the form
\eq{
\dfrac{d\xi}{d\t}=8m_{21}^2\xi_+\xi_-\lb(1-\frac{\xi}{\xi_-}\rb)
\lb(1-\frac{\xi}{\xi_+}\rb),
\label{2.32}
}
where $\xi_\pm$ are the solutions of the equation
\eq{
\xi^2-\frac{b-24m_{21}^2\a}{8m_{21}^2}\xi-\frac{8m_{21}^2+\a b+
\a(b-24m_{21}^2)}{8m_{21}^2}=0.
\notag
}
Finally, defining a new variable by
\eq{
\e^2=\frac{\xi}{\xi_-} \notag,
}
we find the standard equation for the Jacobi elliptic functions
\eq{
\lb(\dfrac{d\e}{d\t}\rb)=\frac{8m_{21}^2\xi_+}{4}\lb(1-\e^2\rb)\lb(1-t\e^2\rb),
\label{2.33}
}
where the modulus $t$ is defined by
\eq{
t=\frac{\xi_-}{\xi_+},
\label{2.34}
}
and the solution takes the form
\eq{
\e=sn\lb(\frac{\sqrt{8m_{21}^2\xi_+}}{2}\t,t\rb).
\label{2.35}
}
In the next section we proceed with calculation of the corrections to
the energy for the system we just described.

\sect{Corrections to the classical energy}

In this section we calculate the corrections to the classical energy 
using the approach developed in \cite{min,minzar}. Let us
 consider a circular pulsating string expanding and contracting on
 $S_5$, which has a center of mass that is moving on an $S_3$
 subspace. We will assume that the string is with fixed spatial
 coordinates in $AdS_5$ (exept the time variable), so the relevant metric is:
 \eq{ ds^2=R^2\,(-dt^2+d\gamma^2+\cos^2\gamma\,d\chi^2+\sin^2\gamma
\,d\,\Omega_3^2),\label{1.1b} }
where $d\Omega_3^2\,=\,g_{ij}\,d\varphi^i\,d\varphi^j$ is the
metric on the $S_3$ subspace, i. e.
$g_{ij}=diag(1,\cos^2\varphi^1,\sin^2\varphi^1)$ and
$R^2=2\pi\alpha^\prime\sqrt{\lambda}$. If we identify $t$ with
$\tau$ and use following classical anzatz: 
\eq{\gamma=\gamma(\tau),\,
\chi=\chi(\tau),\,\varphi^i=n^i\sigma+\alpha^i(\tau),\,i=1,2,3,}
the Nambu-Goto action
\eq{S\,=\,-\frac{1}{2\pi\alpha^{\prime}}\,\int\,d\tau\,d\sigma
\sqrt{-det(\partial_{\alpha}X^{\mu}\partial_{\beta}X_{\mu})}}
then reduces to \ml{
S\,=\,-\sqrt{\lambda}\,\int\,d\tau\,d\sigma\,\sin\gamma\\
\times\,\sqrt{g_{ij}n^{i}n^{j}\left(1-\dot{\gamma}^2-
\dot{\chi}^2\cos^2\gamma\right)+\sin^2\gamma\left[\left(g_{ij}n^i
\dot{\alpha}^j\right)^2-\left(g_{ij}\dot{\alpha}^i\dot{\alpha}^j
\right)\left(g_{ij}n^{i}n^{j}\right)\right]}.}
Now we are going to apply the procedure for calculation of the
anomalous dimensions developed in \cite{min}. For this puspose, we
find first the canonical momenta of our system. Straightforward
calculations give for the momenta
 \eq{\Pi_\gamma=\frac{\sqrt{\lambda}\,\sin\gamma\,
(g_{ij}n^{i}n^{j})\,\dot{\gamma}}{\sqrt{g_{ij}n^{i}n^{j}
\left(1-\dot{\gamma}^2-\dot{\chi}^2\cos^2\gamma\right)+\sin^2\gamma
\left[\left(g_{ij}n^i\dot{\alpha}^j\right)^2-\left(g_{ij}\dot{\alpha}^i
\dot{\alpha}^j\right)\left(g_{ij}n^{i}n^{j}\right)\right]}},}
\eq{\Pi_\chi=\frac{\sqrt{\lambda}\,\sin\gamma\,(g_{ij}n^{i}n^{j})\,
\cos^2\gamma\,\dot{\chi}}{\sqrt{g_{ij}n^{i}n^{j}\left(1-\dot{\gamma}^2-
\dot{\chi}^2\cos^2\gamma\right)+\sin^2\gamma\left[\left(g_{ij}n^i
\dot{\alpha}^j\right)^2-\left(g_{ij}\dot{\alpha}^i\dot{\alpha}^j
\right)\left(g_{ij}n^{i}n^{j}\right)\right]}},}
\eq{\Pi_{\alpha^k}=\frac{\sqrt{\lambda}\,\sin^3\gamma\,
[(g_{ij}n^{i}n^{j})\,\dot{\alpha}^s-(g_{ij}n^{i}\dot{\alpha}^{j})
\,n^s]\,g_{sk}}{\sqrt{g_{ij}n^{i}n^{j}\left(1-\dot{\gamma}^2-
\dot{\chi}^2\cos^2\gamma\right)+\sin^2\gamma\left[\left(g_{ij}n^i
\dot{\alpha}^j\right)^2-\left(g_{ij}\dot{\alpha}^i\dot{\alpha}^j
\right)\left(g_{ij}n^{i}n^{j}\right)\right]}}.}
Solving for the derivatives in terms of the canonical momenta and
substituting back into the Hamiltonian, we find
\eq{H\,=\,\sqrt{\Pi^2_{\gamma}+\frac{\Pi^2_{\chi}}{\cos^2\gamma}+
\frac{g_{ij}\Pi^i\Pi^j}{\sin^2\gamma}+\lambda\,(g_{ij}n^{i}n^{j})
\,\sin^2\gamma},}
or
\eq{H^2\,=\,\Pi^2_{\gamma}+\frac{\Pi^2_{\chi}}{\cos^2\gamma}+
\frac{g_{ij}\Pi^i\Pi^j}{\sin^2\gamma}+\lambda\,(g_{ij}n^{i}n^{j})\,
\sin^2\gamma.}
Since we consider high energies, one can think of this Hamiltonian as
of square root of a point particle one\footnote{When we apply quasiclassical
  quantization we are dealing actually with a family of solutions. Each of them
  looks like ¨almost¨ as for point particle (for more comments see for instance
  \cite{min,minzar}).}. Hence, we can consider the potential terms as
a perturbation. The potential itself has the form
\eq{V(\varphi^1,\gamma)\,=\,\lambda\left[\left((n^1)^2+(n^2)^2\right)+
\left((n^3)^2-(n^2)^2\right)\,\sin^2\varphi^1\right]\,\sin^2\gamma.
\label{0.10}}
 The above perturbation to the free action will produce the
 corrections to the energy and therefore, to the anomalous dimension.
Thus, we proceed with the consideration of the free wave-functions
on $S_5$ and $S_3$, and then use the perturbation theory to first order to
find the correction of order $\lambda$. The total $S_5$ angular
momentum quantum number will be denoted by $L$ and the total
angular momentum quantum number on $S_3$ is $J$.
In these notations we have for $S_3$ and $S_5$ :
 \ml{\triangle(S_3)\,=\,\frac{1}{\sin\varphi^{1}\cos\varphi^{1}}\,
\frac{\partial}{\partial\varphi^1}\left[\sin\varphi^{1}\cos\varphi^{1}
\,\frac{\partial}{\partial\varphi^1}\right]+\\
\frac{1}{\cos^2\varphi^1}\,
\frac{\partial^2}{\partial(\varphi^2)^2}+\frac{1}{\sin^2\varphi^1}\,
\frac{\partial^2}{\partial(\varphi^3)^2},\label{0.11}}\\[8pt]
\eq{\triangle(S_5)\,=\,\frac{1}{\sin^3\gamma\cos\gamma}\,
\frac{\partial}{\partial\gamma}\left[\sin^3\gamma\cos\gamma\,
\frac{\partial}{\partial\gamma}\right]+\frac{1}{\cos^2\gamma}\,
\frac{\partial^2}{\partial\chi^2}+\frac{1}{\sin^2\gamma}\,
\triangle(S_3).\label{0.12}}
Next step is to find the wave function for our system. For this
 purpose we consider the Schrodinger equations which take for three-sphere the form:
\ml{
 \Delta_{S^3}=\left[\frac{1}{\cos\varphi^1\,\sin\varphi^1}\,\frac{d}{d\varphi^1}\,
\left(\cos\varphi^1\,\sin\varphi^1\,\frac{d}{d\varphi^1}\right)-\right.\notag \\
\left. \frac{m^2}{\cos^2\varphi^1}-\frac{l^2}{\sin^2\varphi^1}+J(J+2)\right] \notag}
\eq{ \Delta_{S^5}(\varphi^1) U(\varphi^1)\,=\,0,
\label{0.13}}
and for five-sphere correspondingly
\ml{
\Delta_{S^5}=\left[\frac{1}{\cos\gamma\,\sin^3\gamma}\,\frac{d}{d\gamma}\,
\left(\cos\gamma\,\sin^3\gamma\,\frac{d}{d\gamma}\right)-\frac{M^2}
{\cos^2\gamma}-\right.\\
\left.\frac{J(J+2)}{\sin^2\gamma}+L(L+4)\right]\notag
}
\eq{
 \Delta_{S^3}(\gamma)U(\gamma)\,=\,0.
\label{0.14}}
The solutions to these equations are well known and they
are:
\eq{
U^J_{m,l}(\varphi^1)\,=\,\tan^l\varphi^1\,\cos^J\varphi^1\,F
\left[\frac{l-J+m}{2}\,,\,\frac{l-J-m}{2}\,;\,l+1\,;\,-\tan^2
\varphi^1\right],\label{0.15}}\\[8pt]
\eq{U^L_{M,J}(\gamma)\,=\,\tan^J\gamma\,\cos^L\gamma\,F
\left[\frac{J-L+M}{2}\,,\,\frac{J-L-M}{2}\,;\,J+2\,;\,-\tan^2
\gamma\right].\label{0.16}}\\[5pt]
In addition we have to ensure that the solutions
 $U^J_{m,l}(\varphi^1)$ 
and $U^L_{M,J}(\gamma)$ are square integrable with respect to the measures
 $d\mu(\varphi^1)\,=\,\cos\varphi^1\sin\varphi^1\,d\varphi^1$ and
 $d\mu(\gamma)\,=\,\cos\gamma\sin^3\gamma\,d\gamma$ with the following
 restrictions on the parameters
 \eq{|m|+l=J-2s,\,\,s=0,1,2,...,[\frac{J}{2}],\label{0.17}}
\eq{|M|+J=L-2p,\,\,p=0,1,2,...,[\frac{L}{2}],\label{0.18}} and
$J>0$, $L>0$, $l>0$.
Introducing new variable  $\sin^2\varphi^1\,=\,\omega$,
 $k\,=\,\frac{1}{2}\,(|m|-m+2s)$ (hence $k\,\in\,\mathbb{N}$), we
 have:
 \ml{U^J_{m,l}(\varphi^1)\,=\,\tan^l\varphi^1\,\cos^J\varphi^1\,F
\left[\frac{l-J+m}{2}\,,\,\frac{l-J-m}{2}\,;\,l+1\,;\,-\tan^2
\varphi^1\right]=\\
 =\,\tan^l\varphi^1\,\cos^J\varphi^1\,F\left[\frac{m-|m|-2s}{2}\,,
\,\frac{-m-|m|-2s}{2}\,;\,l+1\,;\,-\tan^2\varphi^1\right]=\\
 =\,\omega^{\frac{l}{2}}\,\left(1-\omega\right)^{\frac{|m|+2s}{2}}\,
F\left[-k\,,\,-m-k\,;\,l+1\,;\,\frac{\omega}{\omega-1}\right]=\\
 =\,\omega^{\frac{l}{2}}\,\left(1-\omega\right)^{\frac{|m|+2s}{2}}\,
\left(1-\omega\right)^{-k}\,F\left[-k\,,\,m+k+l+1\,;\,l+1\,;\,\omega\right]=\\
 =\,\omega^{\frac{l}{2}}\,\left(1-\omega\right)^{\frac{m}{2}}\,
F\left[-k\,,\,m+k+l+1\,;\,l+1\,;\,\omega\right]=\\
 =\,\frac{k!\,\Gamma(1+l)}{\Gamma(k+1+l)}\,\omega^{\frac{l}{2}}\,
\left(1-\omega\right)^{\frac{m}{2}}\,P_{k}^{(l,m)}(1-2\omega),
\label{0.19}}
In this notations the measure for $\phi_1\in S^3$ becomes
\eq{d\mu(\varphi^1)\,=\,\cos\varphi^1\sin\varphi^1\,d\varphi^1\,=\,
\frac{1}{2}\,d\omega\,.\label{20}}
One can easily find the normalization of our wave function computing
 \ml{\langle\,U^J_{m,l}\,|\,U^J_{m,l}\,\rangle\,=\,
\int\limits_{0}^{\frac{\pi}{2}}\,{U^J_{m,l}}^2(\varphi^1)\,d\mu(\varphi^1)\,=\\
 =\,\left[\frac{k!\,\Gamma(1+l)}{\Gamma(k+1+l)}\right]^2\,
\frac{1}{2}\int\limits_{0}^{1}\,\omega^l\,\left(1-\omega\right)^m\,
\left[P_{k}^{(l,m)}(1-2\omega)\right]^2\,d\omega=\\
 =\,\frac{1}{2^{l+m+2}}\,\left[\frac{k!\,\Gamma(1+l)}{\Gamma(k+1+l)}
\right]^2\,\int\limits_{-1}^{1}\,\left(1-x\right)^l\,\left(1+x\right)^m\,
\left[P_{k}^{(l,m)}(x)\right]^2\,dx\,=\\
 =\,\frac{1}{2(2k+l+m+1)}\,\left[\frac{k!\,\Gamma(1+l)}{\Gamma(k+1+l)}
\right]^2\,\frac{\Gamma(k+l+1)\,\Gamma(k+m+1)}{k!\,\Gamma(k+l+m+1)}.\label{0.21}}
Therefore, the normalized wave-function of \eqref{0.15} becomes:
\eq{\Psi^J_{m,l}(\omega)\,=\sqrt{\frac{2(2k+l+m+1)\,k!\,\Gamma(k+l+m+1)}
{\Gamma(k+l+1)\,\Gamma(k+m+1)}}\,\,\omega^{\frac{l}{2}}\,
\left(1-\omega\right)^{\frac{m}{2}}\,P_{k}^{(l,m)}(1-2\omega).\label{0.22}}\\[10pt]
If we introduce $\sin^2\gamma=z$, $ \alpha=J+1$, $\beta=M$,
$n=\frac{1}{2}\,(|M|-m+2p)$ ($n\,\in\,\mathbb{N}$), the measure for
the rest of gemetry becomes
\eq{
d\mu(\gamma)\,=\,\cos\gamma\sin^3\gamma\,d\gamma\,=
\,\frac{1}{2}\,z\,dz\notag
}
and the normalized wave-function of \eqref{0.16} takes the form
\eq{\Psi^L_{M,J}(z)\,=\sqrt{\frac{2(2n+\alpha+\beta+1)\,n!\,
\Gamma(n+\alpha+\beta+1)}{\Gamma(n+\alpha+1)\,\Gamma(n+\beta+1)}}\,
\,z^{\frac{\alpha-1}{2}}\,\left(1-z\right)^{\frac{\beta}{2}}\,
P_{n}^{(\alpha,\beta)}(1-2z).\label{0.23}}\\[10pt]
The potential should be also rewritten in terms of the new
variables. We find for it
\ml{V(\varphi^1,\gamma)\,=\,\lambda\left[\left((n^1)^2+(n^2)^2
\right)+\left((n^3)^2-(n^2)^2\right)\,\sin^2\varphi^1\right]\,\sin^2\gamma=\\
=\,\lambda\left[\left((n^1)^2+(n^2)^2\right)+\left((n^3)^2-(n^2)^2
\right)\,\omega\right]\,z\,=\,V(\omega,z).\label{0.24}}

Now we can compute the corrections to the energy using the
 perturbation theory. The final result for the first order correction to $E^2$ is
 \ml{E^2_{(1)}\,=\,\lambda\,\int\limits_0^{1}\,d\mu(\omega)\,
\int\limits_0^1\,d\mu(z)\,V(\omega,z)\,{\Psi^J}^2_{m,l}(\omega)\,
{\Psi^L}^2_{M,J}(z)\\
 =
 \lambda\,\frac{1}{4}\,\int\limits_0^1\left[\left((n^1)^2+(n^2)^2
\right)+\left((n^3)^2-(n^2)^2\right)\,\omega\right]\,{\Psi^J}^2_{m,l}
(\omega)\,d\omega\,\,\int\limits_0^1\,z^2\,{\Psi^L}^2_{M,J}(z)\,dz\\
 =\,\lambda\,\left[\left((n^1)^2+(n^2)^2\right)+\frac{\left((n^3)^2-(n^2)^2
\right)}{(2k+l+m+1)}\,\left(\frac{(k+l)\,(k+l+m)}{(2k+l+m)}\right.\right.
\\
\left.\left.+ \frac{(k+1)\,(k+m+1)}{(2k+l+m+2)}\right)\right]
\frac{1}{(2n+\alpha+\beta+1)}\times \\
\times\left[\frac{(n+\alpha)\,
(n+\alpha+\beta)}{(2n+\alpha+\beta)}+\frac{(n+1)\,(n+\beta+1)}
{(2n+\alpha+\beta+2)}\right].\label{0.25}}

A simple comparison of our result with that in \cite{minzar} shows
that the results differ significantly. This is because the string
ansazt for the classical solutions in our case is different and
therefore the solutions are actually new and different. The
semicassical quantization we performed means summation over the
classical solutions and therefore the final results should be
different as actually is. 

\sect{Conclusion}

In this section we summarize the results of our study.
The goal we pursued in this paper was to look for more general pulsating string
solutions in $AdS_5\times S^5$ background. Using the idea for
generalized string ansatz suggested in \cite{tsnr}
for folded string, we find new pulsating string solutions. In this case the string
sigma model can be written as constrained system with the help of
Lagrange multiplier $\Lambda$. We find two classes of solutions - for
constant $\Lambda$ and for non-constant $\Lambda$. While in the first
case we obtained fairly general solution, in the latter one we
restricted ourself to the case of $\a_3=v_3=r_3=0$. The corresponding
solutions are found to be simple harmonic functions of $\tau$
($\Lambda=const$) and Jacobi elliptic functions (non-constant
$\Lambda$). Next we considered the corrections to the classical
energy. From AdS/CFT point of view the corrections to the classical
energy gives the anomalous dimensions of the operators in SYM theory
and therefore they are of primary interest. Since we found solutions
for particular ansatz for the calssical string solutions, to obtain
the energy corrections we used the more general approach suggested by
Minahan \cite{min}. For this purpose we consider the Nambu-Goto actions and find
the Hamiltonian. After that we quantize the resulting theory
semiclassically and obtain the corrections to the energy. Since we
consider highly excited system, the kinetic term is dominating. All this means
that we effectively perform summation over all classical solutions
(not only those we explicitly found) while
the effective potential term serves for a small perturbation. Since
our ansatz, and therefore the solutions also, differs from that in
\cite{minzar}, the final result for the corrections to the classical
energy differs significantly.

As a final comment we note that to comlete the analysis from AdS/CFT
point of view, it is of great interest to develop Bethe ansatz
analysis and to compare our result to that in SYM side. We leave this
important question for future research.

\vspace*{.8cm}

{{\large{\bf Acknowledgements:}} R.R. would like to thank A. Tseytlin
  for comments, email correspondence and critically reading the draft
  of the paper.

\end{document}